\documentclass{aa} 
\usepackage{graphicx}
\usepackage{txfonts}
\usepackage{natbib}
\usepackage{url}
\usepackage[usenames]{color}
\bibpunct{(}{)}{;}{a}{}{,} 
\newcount\longrefs
\def\aap{\ifnum\longrefs=1 {Astron.\ Astrophys.}\else 
                           {A\hbox{\rm \&}A}\fi}
\def\aapr{\ifnum\longrefs=1 {Astron.\ Astrophys.\ Rev.}\else 
                            {A\hbox{\rm \&}AR}\fi}
\def\aaps{\ifnum\longrefs=1 {Astron.\ Astrophys.\ Suppl.}\else 
                            {A\hbox{\rm \&}A Suppl.}\fi}
\def\aj{\ifnum\longrefs=1 {Astron.\ J.}\else 
                          {AJ}\fi} 
\def\ao{\ifnum\longrefs=1 {Applied Optics}\else 
                           {Appl.\ Opt.}\fi} 
\def\aspcs{\ifnum\longrefs=1 {Astron.\ Soc.\ Pacific Conf. Series}\else 
                           {ASP Conf.\ Ser.}\fi} 
\def\apj{\ifnum\longrefs=1 {Astrophys.\ J.}\else 
                           {ApJ}\fi} 
\def\apjl{\ifnum\longrefs=1 {Astrophys.\ J. Lett.}\else 
                            {ApJ}\fi} 
\def\aplett{\ifnum\longrefs=1 {Astrophys.\ J. Lett.}\else 
                            {ApJ}\fi} 
\def\apjs{\ifnum\longrefs=1 {Astrophys.\ J. Suppl.}\else 
                            {ApJS}\fi}
\def\apss{\ifnum\longrefs=1 {Astrophys.\ and Space Science}\else 
                            {Astrophys.\ Space Sci.}\fi}
\def\araa{\ifnum\longrefs=1 {Ann.\ Rev.\ Astron.\ Astrophys.}\else 
                            {ARA\hbox{\rm \&}A}\fi}
\def\azh{\ifnum\longrefs=1 {Astronomicheskii Zhurnal}\else 
                            {Astron.\ Zhur.}\fi}
\def\baas{\ifnum\longrefs=1 {Bull.\ Am.\ Astron.\ Soc.}\else 
                            {BAAS}\fi}
\def\bain{\ifnum\longrefs=1 {Bull.\ Astronom.\ Institutes Netherlands}\else
                            {Bull.\ Astr.\ Inst.\ Neth.}\fi}
\def\gca{\ifnum\longrefs=1 {Geochim.\ Cosmochim.\ Acta}\else 
                           {Geochim.\ Cosmochim.\ Acta}\fi}
\def\grl{\ifnum\longrefs=1 {Geophys.\ Res.\ Lett.}\else 
                           {Geoph.\ Res.\ Lett.}\fi}
\def\iaucirc{\ifnum\longrefs=1 {IAU Circulars}\else 
                          {IAU Circ.}\fi}
\def\ip{\ifnum\longrefs=1 {in press}\else 
                          {in press}\fi}
\def\jgr{\ifnum\longrefs=1 {J.\ Geophys.\ Res.}\else 
                           {J.\ Geophys.\ Res.}\fi}  
\def\jrasc{\ifnum\longrefs=1 {J.\ Royal Astron.\ Soc.\ Canada}\else 
                           {JRAS Can.}\fi}  
\def\mnras{\ifnum\longrefs=1 {Mon.\ Not.\ Roy.\ Astron.\ Soc.}\else 
                             {MNRAS}\fi} 
\def\nat{\ifnum\longrefs=1 {Nature}\else 
                           {Nat}\fi}
\def\pasj{\ifnum\longrefs=1 {Pub.\ Astron.\ Soc.\ Japan}\else 
                            {PASJ}\fi} 
\def\pasp{\ifnum\longrefs=1 {Pub.\ Astron.\ Soc.\ Pacific}\else 
                            {PASP}\fi} 
\def\physscr{\ifnum\longrefs=1 {Physica Scripta}\else 
                            {Phys.\ Scrip.}\fi} 
\def\planss{\ifnum\longrefs=1 {Planetary \& Space Science}\else 
                            {Plan. \& Space Sci.}\fi} 
\def\procspie{\ifnum\longrefs=1 {Proc.\ SPIE}\else 
                            {Proc.\ SPIE}\fi} 
\def\qjras{\ifnum\longrefs=1 {Quarterly J.\ Royal Astron.\ Soc.}\else 
                            {QJRAS}\fi} 
\def\sa{\ifnum\longrefs=1 {Soviet Astron..}\else 
                               {Sov.\ Astron.}\fi}
\def\skytel{\ifnum\longrefs=1 {Sky \& Telescope}\else 
                            {Sky \& Tel.}\fi} 
\def\solphys{\ifnum\longrefs=1 {Solar Phys.}\else 
                               {Sol.\ Phys.}\fi}
\def\ssr{\ifnum\longrefs=1 {Space Science Rev.}\else 
                               {Space\ Sci.\ Rev.}\fi}

\def\nl{,\ } 

\def\CMAO{Center of Mathematics for Applications, University of Oslo\nl
           P.O. Box 1053, Blindern\nl N-0316 Oslo\nl Norway}

\def\Oslo{Institute of Theoretical Astrophysics, University of Oslo\nl 
         P.O. Box 1029, Blindern\nl N--0315 Oslo\nl Norway}

\def\SIU{Sterrekundig Instituut, Utrecht University\nl Postbus 80\,000\nl
         NL--3508 TA Utrecht\nl The Netherlands}

\def\dutch{\def\refname{Referenties}\def\abstractname{Samenvatting}%
  \def\bibname{Bibliografie}\def\chaptername{Hoofdstuk}%
  \def\appendixname{Bijlage}\def\contentsname{Inhoudsopgave}%
  \def\listfigurename{Lijst van figuren}%
  \def\listtablename{Lijst van tabellen}%
  \def\indexname{Index}\def\figurename{Figuur}\def\tablename{Tabel}%
  \def\partname{Deel}\def\enclname{Bijlage(n)}\def\ccname{Ter attentie van}%
  \def\headtoname{Aan}\def\headpagename{Pagina}%
  \def\today{\number\day\space\ifcase\month\or januari\or februari\or%
     maart\or%
     april\or mei\or juni\or juli\or augustus\or september\or oktober\or%
     november\or december\fi \space\number\year}%
  \typeout{
              >>>>> use hlatex209 for Dutch hyphenation <<<<< 
         }}

\DeclareFontFamily{OT1}{mvs}{}
\DeclareFontShape{OT1}{mvs}{m}{n}{<-> fmvr8x}{}

\newcounter{onefig} \newcounter{fignumber}
\newcount\nocaptions \newcount\nofigures \newcount\figwidth
\newcount\viewgraphs \newcount\printlabel
\long\def\skipfigure #1\viewout{}   
  \def\paper{}  \def\figlabel{} 
\long\def\nextfig#1{\setcounter{figure}{\value{fignumber}}
  \addtocounter{fignumber}{1}
  \ifnum \viewgraphs=1 \pagestyle{empty} \fi 
  \ifnum\value{onefig}=0 #1 \fi                 
  \ifnum\value{onefig}=\value{fignumber} #1 \fi}
\def\figwidths#1#2{\ifnum \nocaptions=1 #2mm \else #1mm \fi}  
\def\picplace{\framebox[80mm]{\rule{0cm}{1cm}}}
\def\paper#1{}  
\long\def\plotfig#1#2{\ifnum \nofigures=1 \picplace \else #2 \fi}
\long\def\captiontext#1{\ifnum \nofigures=1 \raggedright \fi 
   \ifnum \nocaptions=1 \paper
     \ifnum \viewgraphs=0 
       \newline  \mbox{}\hrulefill\mbox{} \newline 
       \ifnum \printlabel=1 \{{\em \figlabel}\}\newline \fi
     \fi 
   \else \ifnum \printlabel=1 \{{\em \figlabel}\}\newline \fi
     #1 \fi}

\newcount\panelwidth \newcount\panelheight 
\newcount\bxmin \newcount\bymin \newcount\bxmax \newcount\bymax
\newcount\tbxmin \newcount\tbymin
\newcount\tpanelwidth \newcount\tpanelheight \newcount\tpdif
\def\panelsize #1,#2;{\panelwidth=#1 \panelheight=#2}  
\def\setbb #1,#2;#3,#4;#5,#6;{
  \tbxmin=#1 \tbymin=#2    
  \bxmin=#3 \bymin=#4      
  \bxmax=#5 \bymax=#6}     
\def\barepanel #1{%
  \ifnum\panelheight=0 
    \tpdif=\bymax \advance\tpdif by -\bymin
    \multiply \tpdif by \panelwidth
    \tpanelheight=\tpdif
    \tpdif=\bxmax \advance\tpdif by -\bxmin
    \divide \tpanelheight by \tpdif
  \else \tpanelheight=\panelheight \fi
  \ifnum\panelwidth=0 
    \tpdif=\bxmax \advance\tpdif by -\bxmin
    \multiply \tpdif by \panelheight
    \tpanelwidth=\tpdif
    \tpdif=\bymax \advance\tpdif by -\bymin
    \divide \tpanelwidth by \tpdif
  \else \tpanelwidth=\panelwidth \fi
  \epsfig{file=#1,silent=,%
     bbllx=\bxmin bp,bblly=\bymin bp,bburx=\bxmax bp,bbury=\bymax bp,clip=,%
     width=\tpanelwidth mm,height=\tpanelheight mm}}
\def\labelypanel #1{
  \ifnum\panelheight=0 
    \tpdif=\bymax \advance\tpdif by -\bymin
    \multiply \tpdif by \panelwidth
    \tpanelheight=\tpdif
    \tpdif=\bxmax \advance\tpdif by -\bxmin
    \divide \tpanelheight by \tpdif
  \else \tpanelheight=\panelheight \fi
  \ifnum\panelwidth=0 
    \tpdif=\bxmax \advance\tpdif by -\bxmin
    \multiply \tpdif by \panelheight
    \tpanelwidth=\tpdif
    \tpdif=\bymax \advance\tpdif by -\bymin
    \divide \tpanelwidth by \tpdif
  \else \tpanelwidth=\panelwidth \fi
  \tpdif=\bxmax \advance\tpdif by -\tbxmin
  \multiply \tpanelwidth by \tpdif
  \tpdif=\bxmax \advance\tpdif by -\bxmin
  \divide \tpanelwidth by \tpdif
  \epsfig{file=#1,silent=,%
    bbllx=\tbxmin bp,bblly=\bymin bp,bburx=\bxmax bp,bbury=\bymax bp,%
    clip=,width=\tpanelwidth mm,height=\tpanelheight mm}}
\def\labelxpanel #1{%
  \ifnum\panelheight=0 
    \tpdif=\bymax \advance\tpdif by -\bymin
    \multiply \tpdif by \panelwidth
    \tpanelheight=\tpdif
    \tpdif=\bxmax \advance\tpdif by -\bxmin
    \divide \tpanelheight by \tpdif
  \else \tpanelheight=\panelheight \fi
  \ifnum\panelwidth=0 
    \tpdif=\bxmax \advance\tpdif by -\bxmin
    \multiply \tpdif by \panelheight
    \tpanelwidth=\tpdif
    \tpdif=\bymax \advance\tpdif by -\bymin
    \divide \tpanelwidth by \tpdif
  \else \tpanelwidth=\panelwidth \fi
  \tpdif=\bymax \advance\tpdif by -\tbymin
  \multiply \tpanelheight by \tpdif
  \tpdif=\bymax \advance\tpdif by -\bymin
  \divide \tpanelheight by \tpdif
  \epsfig{file=#1,silent=,%
    bbllx=\bxmin bp,bblly=\tbymin bp,bburx=\bxmax bp,bbury=\bymax bp,%
    clip=,width=\tpanelwidth mm,height=\tpanelheight mm}}
\def\labelxypanel #1{%
  \ifnum\panelheight=0 
    \tpdif=\bymax \advance\tpdif by -\bymin
    \multiply \tpdif by \panelwidth
    \tpanelheight=\tpdif
    \tpdif=\bxmax \advance\tpdif by -\bxmin
    \divide \tpanelheight by \tpdif
  \else \tpanelheight=\panelheight \fi
  \ifnum\panelwidth=0 
    \tpdif=\bxmax \advance\tpdif by -\bxmin
    \multiply \tpdif by \panelheight
    \tpanelwidth=\tpdif
    \tpdif=\bymax \advance\tpdif by -\bymin
    \divide \tpanelwidth by \tpdif
  \else \tpanelwidth=\panelwidth \fi
  \tpdif=\bxmax \advance\tpdif by -\tbxmin
  \multiply \tpanelwidth by \tpdif
  \tpdif=\bxmax \advance\tpdif by -\bxmin
  \divide \tpanelwidth by \tpdif 
  \tpdif=\bymax \advance\tpdif by -\tbymin 
  \multiply \tpanelheight by \tpdif
  \tpdif=\bymax \advance\tpdif by -\bymin
  \divide \tpanelheight by \tpdif
  \epsfig{file=#1,silent=,%
    bbllx=\tbxmin bp,bblly=\tbymin bp,bburx=\bxmax bp,bbury=\bymax bp,%
    clip=,width=\tpanelwidth mm,height=\tpanelheight mm}}

\def\CC{\par \vspace*{-2ex} \footnotesize \baselineskip=8pt \begin{verbatim}}

\long\def\startignore #1\stopignore{}   

\def\setlistparams{         
  \topsep=0.7ex                 
  \itemsep=0.7ex                
  \leftmargini=3ex}             
\setlistparams                  

\newcounter{alistindex}       

\newcounter{romenumnr}

\newlength{\minipagewidth}

\newsavebox{\boxcontent}
\newcommand{\ovalhead}[1]{
  \unitlength=1cm
  \sbox{\boxcontent}{\mbox{~~{#1}~~}}
  \begin{center}
    \ifdim\wd\boxcontent>6ex 
    \ifdim\wd\boxcontent<8cm 
    \begin{picture}(8,3) \thicklines     
      \put(4.0,0.8){\oval(8,1.6)} 
      \put(0.0,0.7){\parbox{8cm}{
         \begin{center} \usebox{\boxcontent} \end{center}}}
    \end{picture}
    \else \ifdim\wd\boxcontent<12cm 
    \begin{picture}(12,3) \thicklines     
        \put(6.0,0.8){\oval(12,1.6)} 
        \put(0.0,0.7){\parbox{12cm}{
           \begin{center} \usebox{\boxcontent} \end{center}}}
    \end{picture}
    \else
    \begin{picture}(16,3) \thicklines     
        \put(8.0,0.8){\oval(16,1.6)} 
        \put(0.0,0.7){\parbox{16cm}{
           \begin{center} \usebox{\boxcontent} \end{center}}}
    \end{picture}
    \fi \fi \fi
  \end{center}} 

\newcounter{headnr}            
\newcounter{subheadnr}[headnr]
\newcounter{subsubheadnr}[subheadnr]

\font\dropfont= cmr12 scaled \magstep5
\def\dropcap#1#2{{\noindent
    \setbox0\hbox{\dropfont #1}\setbox1\hbox{#2}\setbox2\hbox{(}%
    \count0=\ht0\advance\count0 by\dp0\count1\baselineskip
    \advance\count0 by-\ht1\advance\count0by\ht2
    \dimen1=.5ex\advance\count0by\dimen1\divide\count0 by\count1
    \advance\count0 by1\dimen0\wd0
    \advance\dimen0 by.25em\dimen1=\ht0\advance\dimen1 by-\ht1
    \global\hangindent\dimen0\global\hangafter-\count0
    \hskip-\dimen0\setbox0\hbox to\dimen0{\raise-\dimen1\box0\hss}%
    \dp0=0in\ht0=0in\box0}#2}

\def\rmit#1{{\it #1}}              
           
\def\ie{\rmit{i.e.,}}              

\def\specchar#1{\uppercase{#1}}    

\def\CaII{\mbox{Ca\,\specchar{ii}}}

\def\Halpha{\mbox{H\hspace{0.1ex}$\alpha$}} 

\def\level #1 #2#3#4{$#1 \: ^{#2} \mbox{#3} ^{#4}$}   

\def\rma{{\rm a}}              
  
\def\rmd{{\rm d}}  
\def\rme{{\rm e}}

\def\rml{{\rm l}}

\def\rmr{{\rm r}}

\def\rmu{{\rm u}}

\def\={\hbox{$\!=\!$}}                     

\def\mathstacksym#1#2#3#4#5{\def#1{\mathrel{\hbox to 0pt{\lower 
    #5\hbox{#3}\hss} \raise #4\hbox{#2}}}}

\mathstacksym\lta{$<$}{$\sim$}{1.5pt}{3.5pt} 
\mathstacksym\gta{$>$}{$\sim$}{1.5pt}{3.5pt} 
\mathstacksym\lrarrow{$\leftarrow$}{$\rightarrow$}{2pt}{1pt} 
\mathstacksym\lessgreat{$>$}{$<$}{3pt}{3pt} 

\def\exp{\rme}
\def\Trad{\ensuremath{T_{\rmr \rma \rmd}}}

\def\is{\ensuremath{\!=\!}}
\def\ntot{\ensuremath{n_\mathrm{H}^{\mathrm{tot}}}}
\def\nother{\ensuremath{n_{\mathrm{noH}}}}
\def\me{\ensuremath{m_{\rme}}}
\def\nel{\ensuremath{n_\rme}}
\def\neother{\ensuremath{n_\rme^{\mathrm{noH}}}}
\def\eother{\ensuremath{e_{\mathrm{noH}}}}
\def\ei{\ensuremath{e_{\mathrm{i}}}}
\def\eh2{\ensuremath{e_{\mathrm{H2}}}}
\def\nh2{\ensuremath{n_{\mathrm{H2}}}}
\def\exp{\rme}
\def\Trad{T_{\rmr \rma \rmd}}

\begin{document}

\title{Non-equilibrium hydrogen ionization in 2D simulations of the
  solar atmosphere.}

\titlerunning{}
  
\subtitle{}

   \author{J.~Leenaarts \inst{1,2}
     \and
     M.~Carlsson \inst{2,3}
     \and
     V.~Hansteen \inst{2,3}
     \and
     R.~J.~Rutten \inst{1,2}
   }

   \offprints{ J. Leenaarts, \\ \email{j.leenaarts@astro.uu.nl} }

   \institute{ \SIU \and \Oslo \and \CMAO}
   
   \date{Received; accepted}

   \abstract
   {The ionization of hydrogen in the solar chromosphere and
     transition region does not obey LTE or instantaneous statistical
     equilibrium because the timescale is long compared with important
     hydrodynamical timescales, especially of magneto-acoustic shocks.
     Since the pressure, temperature, and electron density depend
     sensitively on hydrogen ionization, numerical simulation of the
     solar atmosphere requires non-equilibrium treatment of all
     pertinent hydrogen transitions.  The same holds for any
     diagnostic application employing hydrogen lines.}
   {To demonstrate the importance and to quantify the effects of
     non-equilibrium hydrogen ionization, both on the dynamical
     structure of the solar atmosphere and on hydrogen line formation,
     in particular \Halpha.}
   {We implement an algorithm to compute non-equilibrium hydrogen
    ionization and its coupling into the MHD equations within an
    existing radiation MHD code, and perform a two-dimensional
    simulation of the solar atmosphere from the convection zone to the
    corona.}
  {Analysis of the simulation results and comparison to a companion
    simulation assuming LTE shows that: a) Non-equilibrium computation
    delivers much smaller variations of the chromospheric hydrogen
    ionization than for LTE.  The ionization is smaller within shocks
    but subsequently remains high in the cool intershock phases.  As a
    result, the chromospheric temperature variations are much larger
    than for LTE because in non-equilibrium, hydrogen ionization is a
    less effective internal energy buffer.  The actual shock
    temperatures are therefore higher and the intershock temperatures
    lower.  b) The chromospheric populations of the hydrogen $n \is 2$
    level, which governs the opacity of H$\alpha$, are coupled to the
    ion populations. They are set by the high temperature in shocks and
    subsequently remain high in the cool intershock phases.  c) The
    temperature structure and the hydrogen level populations differ
    much between the chromosphere above photospheric magnetic elements
    and above quiet internetwork.  d) The hydrogen $n \is 2$
    population and column density are persistently high in dynamic
    fibrils, suggesting that these obtain their visibility from being
    optically thick in \Halpha\ also at low temperature.}
  {}
          
  
   \maketitle

\section{Introduction}                          \label{sec:introduction}

The chromosphere represents the least understood regime of the sun
\citep{1998SSRv...85..187J}.
In this paper we address the treatment of hydrogen ionization in
simulations of the solar chromosphere.  It is of paramount importance
because hydrogen makes up 90\% of the nuclei in the solar atmosphere,
is an important and often dominant electron donor, and contains a
large part of the internal energy of the gas. Hence, the ionization
state of hydrogen strongly influences the temperature, pressure and
electron density. Radiation magnetohydrodynamic (MHD) simulations of
the atmosphere must therefore account properly for hydrogen
ionization. This is not only important for the structure of the
atmosphere within a simulation, but also for subsequent computation of
the emergent spectrum from the simulation for comparison with
observations.

\citet{1976ApJ...205..499K,1978ApJ...220.1024K}
and 
\citet{1980A&A....87..229K}
already showed from idealized one-dimensional (1D) models that the
assumption of instantaneous statistical equilibrium (SE) for hydrogen
ionization does not hold in a dynamical atmosphere containing shock
waves.  The temperature difference between the hot shocks and the cool
intershock phases produces disparate ionization and recombination
timescales, the latter being far slower than the former.

  \citet{1992ApJ...397L..59C,1995ApJ...440L..29C,2002ApJ...572..626C} 
computed dynamical 1D simulations of the solar atmosphere including a
detailed non-equilibrium treatment of the hydrogen rate equations
including ionization and recombination (i.e.\ not instantaneous
statistical equilibrium but employing pertinent time derivatives in
the population rate equations, as in Appendix~A of the present paper).
In the first paper they showed that the non-equilibrium effects lead
to significant increase in shock temperature compared with the case of
instantaneous LTE ionization and recombination.  In the second paper
they supply a detailed analysis of the hydrogen ionization in their
simulations.  They found that the chromospheric hydrogen
ionization/recombination timescale is of the order of 50~s within hot
shocks and $10^3 - 10^5$~s in the cool intershock regions, and that
hydrogen becomes partially ionized within shocks but, owing to the
long recombination timescale, does not recombine in the subsequent
post-shock phase.  As a consequence, the degree of ionization of
hydrogen in the chromosphere is rather constant with time, in stark
contrast to what the classical assumptions of statistical equilibrium
or LTE would predict.

The present limitations on computing power do not yet permit such
full-fledged non-equilibrium treatment of hydrogen ionization including
radiative transfer in 2D and 3D simulation geometry.  Therefore,
approximations remain necessary to make the problem computationally
tractable. In this work we employ the method formulated by
 \citet{sollum1999}.

Sollum showed that in 1D dynamical modeling it is possible to avoid
detailed evaluation of the radiation field in the relevant hydrogen
transitions by prescribing suitable radiative rates.  He found that
this approximation works well up to just below the transition
region. Sollum's method is fast enough for use in multi-dimensional
geometry. It was earlier implemented in the code CO$^5$BOLD
\citep{2002AN....323..213F}
to perform a 3D non-magnetic hydrodynamical simulation of the
chromosphere by
 \citet{2006A&A...460..301L}. 
This simulation confirmed the conclusion obtained by
  \citet{1992ApJ...397L..59C, 2002ApJ...572..626C} 
from 1D simulation: the degree of hydrogen ionization in the middle
and upper chromosphere is determined by the passage of
high-temperature shocks, irrespective of the cool intershock phases.
It is relatively constant at about $10^{-3}$ to $10^{-2}$. 
However,
 \citet{2006A&A...460..301L} 
did not implement back-coupling of the non-equilibrium ionization into
the equation of state for the gas within the hydrodynamical
simulation.  This is difficult to do in the CO$^5$BOLD code because it
employs an approximate Riemann solver for the hydrodynamics.

In this paper we first present our implementation of Sollum's method
including back-coupling of the non-equilibrium ionization into the
equation of state in the Oslo MHD code described by
  \citet{2004IAUS..223..385H} 
which uses a finite difference scheme for the hydrodynamics.  We then
discuss an extensive 2D simulation of chromospheric fine-structure
evolution with this code and analyze the results in terms of the
hydrogen ionization balance, with separation between the chromospheric
behavior above a magnetic element and in an area resembling quiet-sun
internetwork.  The effects of non-equilibrium hydrogen ionization are
demonstrated quantitatively through comparison with a similar
companion simulation.  The latter was started with identical initial
conditions but its hydrogen ionization was set to obey LTE, i.e.\
instantaneous Saha--Boltzmann partitioning.

\section{Method}
\label{sec:method}

In this section we describe our implementation of Sollum's method in
the MHD code developed in Oslo
  \citep{2004IAUS..223..385H}. 
%
This code is based on the staggered grid code described in
  \citet{1998A&A...338..329D},  
  \citet{2001SoPh..198..289M}
and by Galsgaard \& Nordlund\footnote{See
  \url{http://www.astro.ku.dk/~kg/}.}.  It includes the multi-group
opacity-binning method of
  \citet{1982A&A...107....1N}
and the treatment of scattering of
  \citet{2000ApJ...536..465S}
for radiative transfer in the photosphere and chromosphere. In the
transition region and corona it employs optically thin radiative
cooling. In addition, it treats radiative losses in strong lines and
continua of hydrogen and \CaII\ in the upper chromosphere and
transition region using an approximation based on detailed 1D
computation with the RADYN code of
  \citet{1992ApJ...397L..59C}. 
Thermal conduction along magnetic field lines is taken into account
  \citep{2004IAUS..223..385H}. 

In this section we first summarize the temporal evolution scheme of
the MHD algorithm and then specify the modifications through which we
insert non-equilibrium hydrogen ionization, the boundary conditions,
and our simulation setup.

\subsection{MHD evolution scheme}

The MHD solver is the explicit third-order predictor-corrector scheme
developed by
 \citet{hyman1979} 
but modified to allow variable timesteps. Its fundamental variables
are the density $\rho$, momentum $\vec{p}$, internal energy density
\ei\ and the magnetic field $\vec{B}$.  The evolution equations for
the fundamental variables depend only on the variables themselves, the
temperature, the gas pressure, and their spatial derivatives. These
derivatives are computed using a sixth-order scheme. The temperature
and pressure are looked up in precomputed tables as function of
density and internal energy. This table is computed assuming LTE
except where the density and internal energy have typical coronal
values. There coronal equilibrium is assumed.

The evolution equation for fundamental variable $f$ is
\begin{equation} \label{eq:predictor}
  \frac{\partial f(t)}{\partial t} = \dot{f}
\end{equation}
where $\dot{f}$ only depends on quantities known at time $t$. Hyman's
scheme solves this equation from timestep $n$ to timestep $n+1$ as
follows: the predictor step is
\begin{equation}
  f^{(*)}_{n+1} = a_1 f_{n-1} + (1-a_1) f_n + b_1 \dot{f}_n,
\end{equation}
and the corrector step 
\begin{equation} \label{eq:corrector}
  f_{n+1} = a_2 f_{n-1} + (1-a_2) f_n + b_2 \dot{f}_n + c_2 \dot{f}^{(*)}_{n+1},
\end{equation}
where $a_1$, $a_2$, $b_1$, $b_2$ and $c_2$ are coefficients that
depend on the current and previous timestep sizes.

\subsection{Implementation of non-equilibrium hydrogen ionization}

In order to compute non-equilibrium hydrogen ionization one has to
solve the hydrogen rate equations
\begin{equation}
 \frac{\partial n_i}{\partial t} + \nabla \cdot (n_i\vec{v}) =
 \sum_{j,j \ne i}^{n_{\rml}} n_j P_{ji} - n_i \sum_{j,j \ne i}^{n_{\rml}} P_{ij},
 \label{eq:RateEq} 
\end{equation}
where $n_i$ is the population of hydrogen level $i$, $\vec{v}$ the
macroscopic velocity, $n_{\rml}$ the number of levels and $P_{ij}$ the
transition rate coefficient between levels $i$ and $j$. The left-hand
side represents a continuity equation for the hydrogen populations,
the right-hand side a source term describing the transitions between
the hydrogen levels. These equations are solved using operator
splitting. The continuity part
\begin{equation}
 \frac{\partial n_i}{\partial t}  = - \nabla \cdot (n_i\vec{v})
\end{equation}
is solved using Hyman's scheme in tandem with the fundamental
variables. It is not possible to use sixth-order spatial derivatives
for the hydrogen populations because negative populations then arise
occasionally from the steep gradients in the population
densities. Instead, a positive-definite first-order upwind scheme is
used to ensure positivity of the populations.

After the predictor step, the rate part of the equations,
\begin{equation}
 \frac{\partial n_i}{\partial t} = \sum_{j,j \ne i}^{n_{\rml}} n_j P_{ji} -
 n_i \sum_{j,j \ne i}^{n_{\rml}} P_{ij},
 \end{equation}
is integrated over the timestep $\Delta t$ while enforcing charge
conservation, hydrogen nuclei conservation, energy conservation and
instantaneous chemical equilibrium between atomic and molecular
hydrogen. This yields predicted values of the hydrogen populations,
temperature and pressure. Subsequently, the corrector step is
performed for the fundamental variables and the advection part of the
hydrogen populations. After this step the rate equations, charge,
energy and particle conservation and chemical equilibrium are solved
again to obtain the hydrogen populations, temperature and pressure for
the new timestep.  The algorithm can be summarized as follows:
\begin{itemize} \vspace{-1ex}
\item predict the fundamental variables and advection of hydrogen
  populations;
\item solve the rate equations and the conservation equations for the
  predicted temperature and pressure;
\item correct the fundamental variables and advection of hydrogen
  populations;
\item solve the rate equations and the conservation equations for the
  corrected hydrogen populations, temperature and pressure.
\end{itemize}

The radiative rate coefficients $R_{ij}$ that enter in the total rate
coefficients $P_{ij}$ are computed using Sollum's method. For each
radiative transition a depth-dependent radiation temperature $\Trad$
is prescribed. It is set equal to the local gas temperature in the
deep layers of the atmosphere, ensuring LTE populations there. Above a
certain height the radiation temperature follows a smooth transition
to a predefined value and then becomes constant at that value. The
latter values were determined by requiring that Sollum's method
reproduces the non-equilibrium hydrogen populations obtained in 1D
modeling of the solar chromosphere with the RADYN code.  All radiative
Lyman transitions are set to obey detailed balancing.  Further detail
is given in
\citet{sollum1999} 
and
\citet{2006A&A...460..301L}. 

The pressure and temperature are explicitly computed in this
modification; the equation of state tables are not used except at the
lower and upper boundaries (Sec.~\ref{sec:bcs}).  The radiative losses
due to hydrogen are still computed time-independently using
Skartlien's multi-group scheme which employs tabulated group-mean
opacities, scattering probabilities and Planck functions based on LTE
populations.

The rate equations, the energy, charge and nucleus conservation
equations and the chemical equilibrium equation together form a set of
coupled non-linear equations that are solved using a Newton-Raphson
scheme. These equations and their derivatives are specified in the
Appendix.

\subsection{Boundary conditions} \label{sec:bcs}

The lower boundary condition enforces LTE hydrogen populations by
solving for Saha-Boltzmann equilibrium instead of the rate equations.
The hydrodynamic conditions regulate the mass flow across the
boundary. Consistency between the total hydrogen number density and
the mass density is automatically enforced through the equation of
hydrogen nucleus conservation. Thus, it is not necessary to specify
hydrogen level population flows across the lower boundary, and the
population continuity equations do not have to be solved in the grid
cells at the lower boundary.

The upper boundary, located in the corona, uses the same approach but
here the non-equilibrium rate equations are solved instead of the
Saha-Boltzmann equations. Consistency between the mass density and the
hydrogen density is enforced by adding or removing protons. This
boundary condition produces coronal equilibrium because the ionization
relaxation timescale of 0.1~s is small compared to the dynamical
timescale in the corona
\citep[see Fig.~6 of][]{2002ApJ...572..626C}.

The equation of state tables are used for the boundary conditions on
the hydrodynamic variables. The tabulated values are accurate at the
boundaries, because they are based on LTE or coronal equilibrium
consistent with the local hydrogen populations.

\subsection{Simulation setup}

We have performed a two-dimensional simulation including
non-equilibrium hydrogen ionization in a setup similar to the one used
by
\citet{2006ApJ...647L..73H}
 and
\citet{2007ApJ...655..624D}.  
It has a horizontal extent of 16.64~Mm with a resolution of 32.5~km
and 512 cells in the horizontal direction. The vertical extent is
11.1~Mm with 150 cells. The vertical resolution varies from 32~km in
the convection zone to 440~km in the corona. Continuum optical depth
unity is located about 1.5~Mm above the lower boundary. The horizontal
boundary condition is periodical. Both the lower and upper boundary
conditions are open, allowing flows to leave and enter the box. The
upper boundary strives to maintain the temperature there at 800,000~K
because the current two-dimensional models do not have sufficient
magnetic dissipation to maintain a corona self-consistently
 \citep{2007ApJ...655..624D}. 
We use a five-level plus continuum hydrogen model atom to compute the
non-equilibrium hydrogen ionization. A heating term is added to the
energy equation that drives the temperature back to 2,400~K when it
falls below that value. This value has no physical meaning but is
present for stability reasons.

The simulation was started from a relaxed snapshot of a previous
simulation which employed LTE ionization and ran for 30~min of solar
time. The effects of the LTE ionization disappeared in approximately
5~min of solar time (see Fig.~1 of
 \citet{2006A&A...460..301L}).
%

\begin{figure*}
  \centering \includegraphics[width=\textwidth]{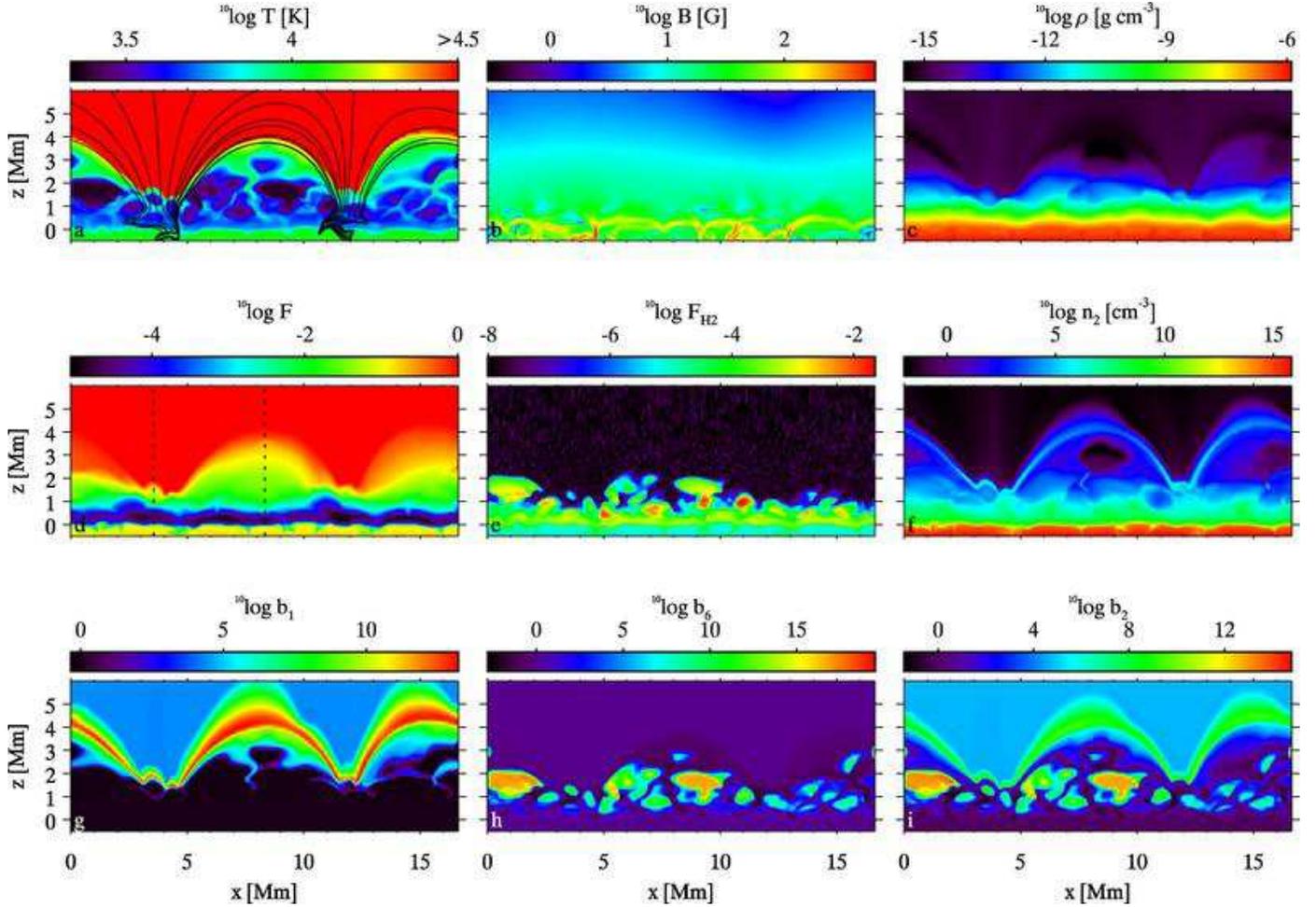}
  \caption{Snapshot cutouts from the simulation, showing various
  quantities in a vertical plane after 8.5~minutes of solar-time
  evolution.  Panel a: gas temperature, with magnetic field lines that
  extend into the corona overplotted in black; b: magnetic field
  strength; c: mass density d: non-equilibrium ionization degree of
  hydrogen; e: fraction of hydrogen atoms in the form of H$_2$
  molecules; f: hydrogen $n \is 2$ level population; g: departure
  coefficient for the hydrogen $n \is 1$ level population; h:
  departure coefficient for the hydrogen $n \is 6$ level population;
  i: departure coefficient for the hydrogen $n \is 2$ level
  population. The columns used in Figs.~\ref{fig:columns} and
  \ref{fig:ztslice} are indicated by dotted lines in panel~d.
  \label{fig:xyslice}}
\end{figure*}

\section{Results}

Figure~\ref{fig:xyslice} shows a snapshot of the simulation. Panel~a
displays the temperature. Some magnetic field lines that extend into
the corona have been overplotted. All coronal field lines are rooted
in two photospheric magnetic field concentrations at $x \is 4$ and $x
\is 11$~Mm. Henceforth we refer to these concentrations as magnetic
elements and to the areas between them as internetwork.

The areas without field lines are not field free, as can be seen in
panel~b which shows the magnitude of the magnetic field strength. The
temperature panel displays granulation at $z \is 0$~Mm, a shock-ridden
chromosphere up to $2-4$~Mm height, and a hot corona reaching peak
temperatures of $10^6$~K. The height and shape of the transition
region strongly depend on the magnetic field configuration, with the
corona reaching deeper down above the magnetic elements. Panel~c shows
the mass density.  It reaches minimum values in the transition region
above internetwork, which consists of extended high-rising arcs (black
in the panel).

Panel~d shows the non-equilibrium hydrogen ionization degree. It has a
minimum between 0~and 0.5~Mm and smoothly increases upward to the
completely ionized corona. The chromospheric ionization degree does
not follow the local gas temperature.  Panel~e shows the fraction of
hydrogen nuclei bound in H$_2$ molecules, which peaks in cool
chromospheric regions between 0.5 and 2~Mm. The black-purple noise
above $z \is 2$~Mm is
a numerical artifact caused by the single precision output of the
code. The code itself uses double precision to avoid such
artifacts. Panel~f shows the population density of hydrogen in the $n
\is 2$ level, the lower level of the \Halpha\ line. It roughly follows
the density structure, with the exception of the transition region
where it shows a sharp increase at the locations of 
the arcs of minimum mass density.

Panel~g shows the non-LTE population departure coefficient of the
ground state of hydrogen $b_1$. The ground state population remains
close to LTE throughout most of the photosphere and chromosphere,
except in strong chromospheric shocks where there is under-ionization
compared to LTE. The ground state is strongly overpopulated in the
transition region and is in coronal equilibrium in the corona.

Panel~h shows the departure coefficient of the hydrogen ion density
$b_6$. It is much larger than unity in chromospheric cool intershock
regions and smaller than unity within chromospheric shocks. This
demonstrates that the non-equilibrium ionization degree is higher than
in LTE in intershock areas and lower in shocks.

The departure coefficient of the $n \is 2$ level in panel~i shows the
same structure in the photoshere and chromosphere as $b_6$ due to the
strong coupling of the continuum and the excited neutral levels. In
the corona, $b_2$ is around $5 \times 10^4$, its coronal equilibrium
value.


\begin{figure*}
 \centering \sidecaption
 \includegraphics[width=12cm]{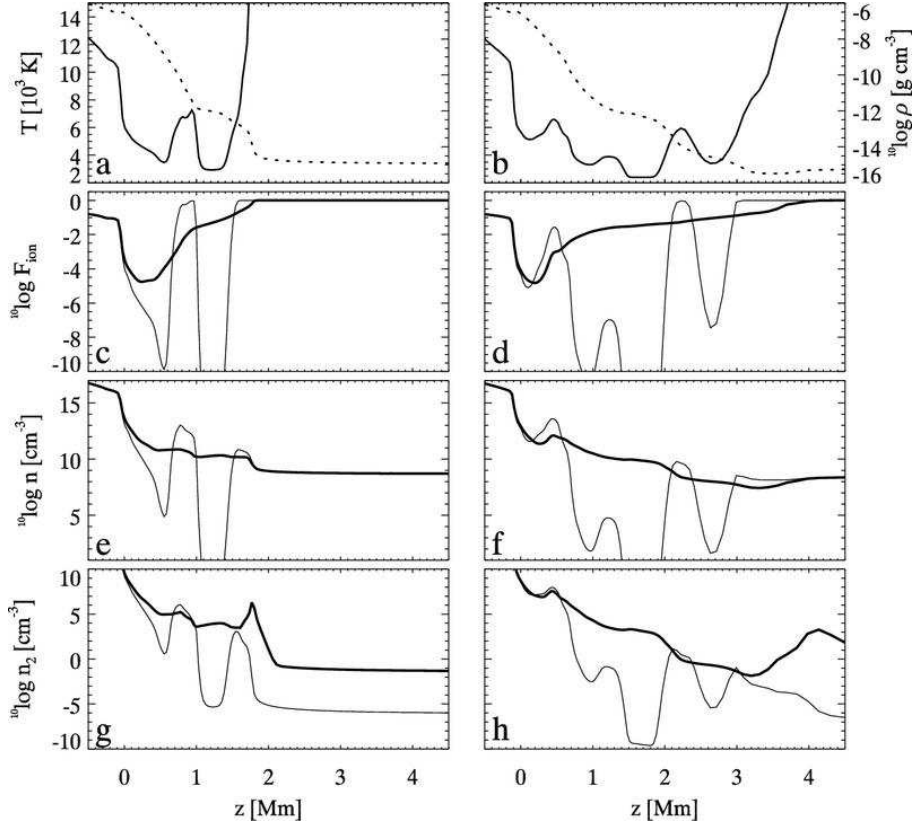}
 \caption{Properties of the simulation along a column in a magnetic
   element (left-hand column) and in the internetwork (right-hand
   column). Panels a and b: temperature (solid) and mass density
   (dashed, right-hand scale); c and d: non-equilibrium (thick) and
   LTE ionization degree (thin); e and f: non-equilibrium (thick) and
   LTE proton density (thin); g and h: population of the $n \is 2$
   level for the non-equilibrium (thick) and LTE (thin) case.
 \label{fig:columns}}
\end{figure*}

\begin{figure*}
  \centering
  \includegraphics[width=\textwidth]{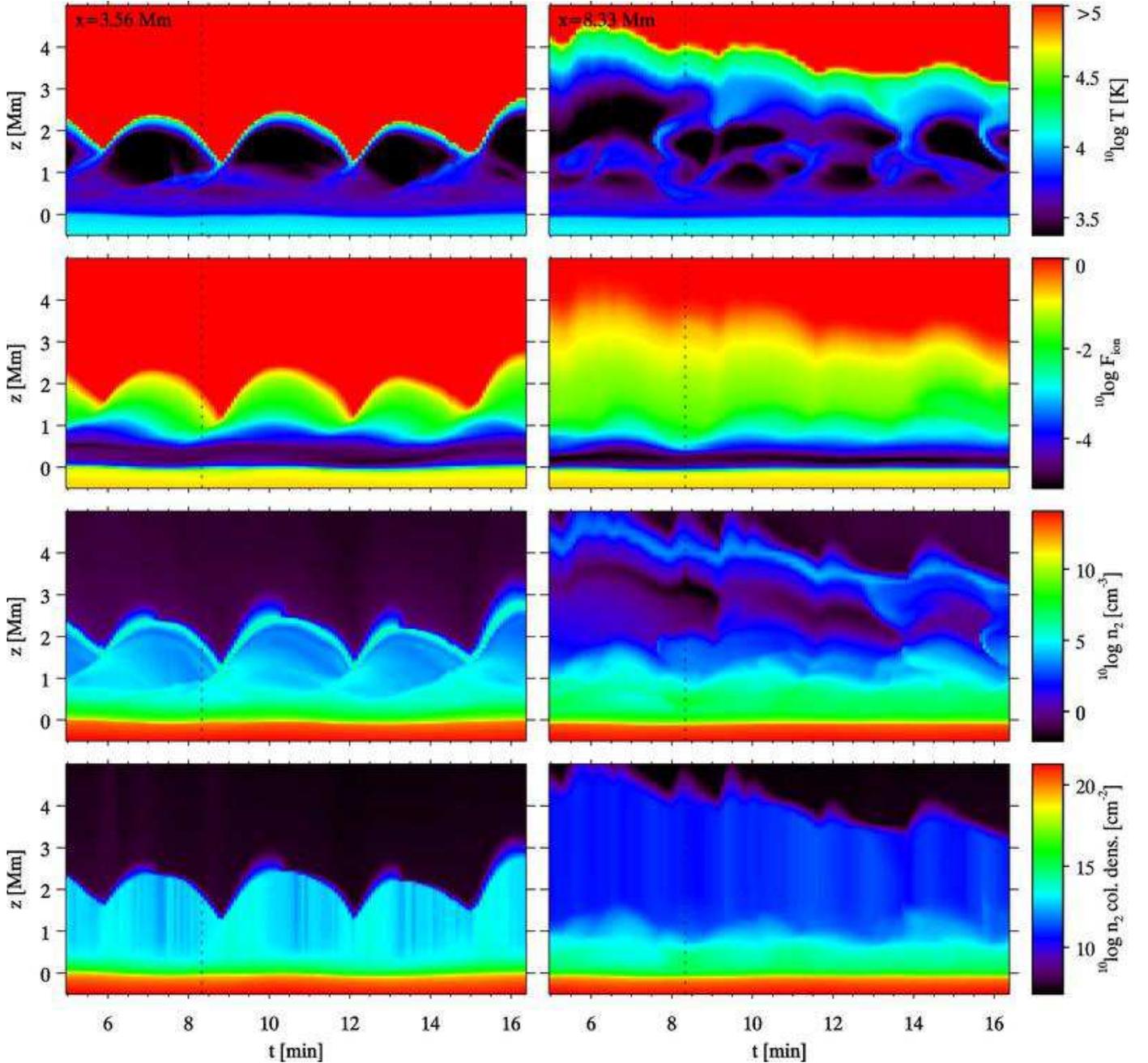}
  \caption{Time slices of the gas temperature (first row), the
    ionization degree of hydrogen (second row), fraction of hydrogen
    in the $n \is 2$ level (third row), and the $n \is 2$ column
    density (fourth row) in a magnetic element (left-hand column) and
    in the internetwork (right-hand column).  The upper-left magnetic
    element panel shows dynamic fibrils pushing the corona upward with
    3~min periodicity. The upper-right internetwork panel shows rather
    unstructured shocks and a slowly varying height of the transition
    region. The snapshot used in Fig.~\ref{fig:xyslice}
    and~\ref{fig:columns} is indicated by a black dotted line.
\label{fig:ztslice}}
\end{figure*}

\begin{figure*}
  \centering
  \includegraphics[width=\textwidth]{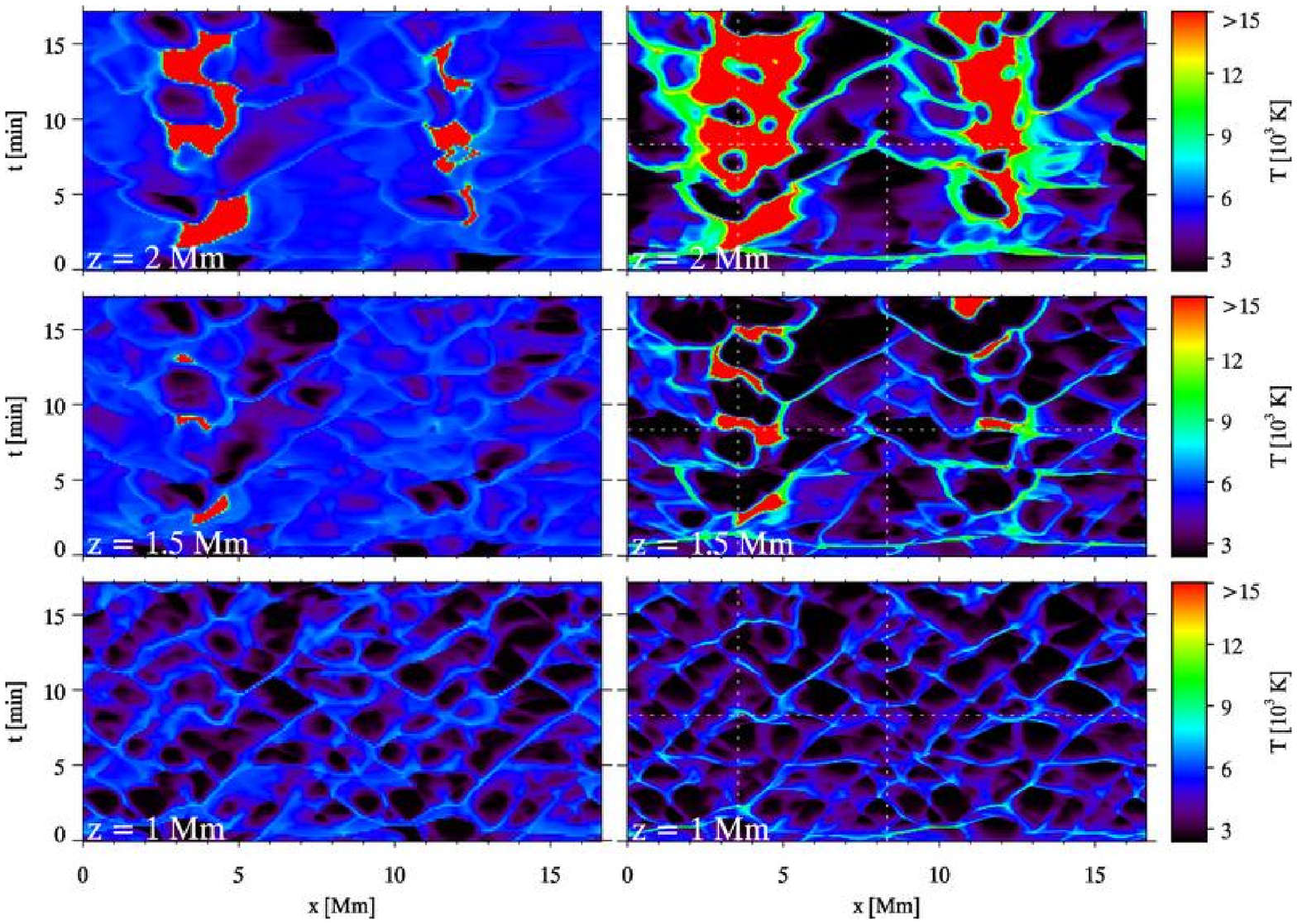}
  \caption{Time slices of the gas temperature at 2, 1.5 and 1~Mm
    height (top to bottom). Left-hand column: simulation with LTE
    ionization; right-hand column: with non-equilibrium
    ionization. Both simulations were started with the same snapshot
    at $t \is 0$~min. The intermittently present red patches are
    caused by the corona, whose lower boundary is moved up and down by
    shocks traveling upward above the magnetic elements. The snapshot
    used in Fig.~\ref{fig:xyslice} is indicated with the horizontal
    dotted line; the columns used in Figs.~\ref{fig:columns} and
    \ref{fig:ztslice} with vertical dotted lines.
  \label{fig:xtslice}}
\end{figure*}

Figure~\ref{fig:columns} shows the behavior of atomic hydrogen along
the two columns marked in Fig~\ref{fig:xyslice}.  These were selected
to sample a magnetic element (left-hand column) and quiet internetwork
(right-hand column). Panels~a and~b show the temperature and mass
density.  The corona starts much lower above the magnetic element than
above the internetwork. Panel~a shows a strong shock at 1~Mm which
will become a dynamic fibril when it reaches the corona, pushing it
upward. In contrast, internetwork panel~b shows no strong shocks. The
density at the transition region is much lower in this case.

Panels~c and~d show the non-equilibrium degree of ionization of
hydrogen as thick curves. It reaches a minimum of $10^{-5}$ at around
0.5~Mm and increases smoothly towards complete ionization in the
corona.  The corresponding LTE ionization, obtained from the
simulation temperature and electron density stratifications with the
Saha-Boltzmann equations, is shown as a thin curve. The dramatic
differences between the curves demonstrate the failure of
instantaneous LTE ionization in the chromosphere and transition
region. In the non-equilibrium case, the slowness of ionization and
recombination prevents total ionization in the shocks and full
recombination in their wakes, producing far smoother ionization
behavior with time than LTE would predict
\citep[see][]{2002ApJ...572..626C,2006A&A...460..301L}.
Note that the LTE curves reach complete ionization in the transition
region at slightly lower heights than the non-equilibrium ionization
curves.  Since hydrogen becomes the dominant electron provider already
at $10^{-4}$ ionization, the electron density equals the proton
density except in the temperature minimum.

Panels~e and~f show the non-equilibrium (thick) and LTE (thin) proton
densities.  Since hydrogen becomes the dominant electron provider
already at $10^{-4}$ ionization, the electron density equals the
proton density except in the ionization minimum around 0.5~Mm
(panels~c and~d).

Panels~g and~h show the population density of the $n \is 2$ level of
hydrogen $n_2$ (thick: non-equilibrium; thin: LTE). It shows the same
trend as the proton population, because all excited hydrogen levels
are strongly coupled to the proton population reservoir. Thus, the
small temporal variation of the ionization is followed by the $n \is
2$ population, and, therefore, the \Halpha\ opacity.  In particular,
the peaks in $n_2$ in the transition region in both panels, at $z \is
1.8$~Mm and $z \is 4.2$~Mm, respectively, provide significant
\Halpha\ visibility. 

Figure~\ref{fig:ztslice} shows the temporal evolution of the
temperature in a magnetic element and in the internetwork. The
magnetic element acts as a wave guide, in which shocks travel upward
with a period of three minutes. When they reach the corona they push
it upward. With time, these motions produce characteristic parabolic
height variation. The same behavior is observed in so-called \Halpha\
dynamic fibrils. The recent papers by
  \citet{2006ApJ...647L..73H}
and
  \citet{2007ApJ...655..624D}
show that \Halpha\ dynamic fibrils are the observational signature of
such magneto-acoustic shocks. Notice that the material in the wake of
the shocks can be as cool as 2,400~K.

The internetwork temperature is less structured. A movie of the gas
temperature (available in the online material for this article) shows
that the internetwork chromosphere is pervaded by shocks which
originate in the photosphere but often travel sideways, away from the
magnetic elements, and sometimes even downward. The transition region
is located higher than above a magnetic element and exhibits
relatively small and irregular temporal variations in height.

The second row shows the ionization degree of hydrogen. It is rather
constant in time in the chromosphere. Above 1~Mm height, it stays at
about 1\% ionization, both in the magnetic element and the
internetwork. The transition to coronal temperatures is smoother in
the internetwork and the increase in ionization is correspondingly
smoother. The third row displays the population of hydrogen in the $n
\is 2$ level $n_2$. It is very low in the almost completely ionized
corona. The transition region shows a local maximum, which is
persistent in time. The $n \is 2$ population is higher in the
transition region of the magnetic element than in the internetwork.
The width of the transition region maximum in the upward phase of the
dynamic fibrils decreases suddenly when the fibril descends again.

The fourth row displays the column density of $n_2$, which is
proportional to the vertical optical depth in the \Halpha\ line. In
the internetwork, the column density starts to increase higher in the
atmosphere than in the magnetic elements, because the chromosphere
extends to larger heights in the internetwork. However, the column
density at the top of the dynamic fibrils (the top of the light blue
bulges) is $10^{13}$~cm$^{-2}$, two orders of magnitude larger than
the internetwork column density at equal height. The same column
density in the internetwork is reached at a height of 0.8~Mm.

\subsection{Comparison with companion LTE simulation}

Figure~\ref{fig:xtslice} presents a comparison between the simulation
with non-equilibrium hydrogen ionization and a simulation with the old
code which employs LTE ionization. Both simulations were started at $t
\is 0$~min from the same relaxed snapshot computed with LTE
ionization.

The bottom panels of~\ref{fig:xtslice} show that at height $z \is
1$~Mm the different treatments of ionization produce only a slight
difference in temperature variation. The wave pattern is almost
identical, but with non-equilibrium ionization the fluctuations are
larger. At larger heights, the differences between the simulations are
large (upper two rows). The shock patterns (thin hot threads) differ
markedly. The shock temperatures are much larger in the non-equilibrium
simulation. They typically are 10,000~K and sometimes reach 13,000~K,
whereas the LTE shock temperatures do not exceed 8,000~K. These high
temperatures are due to the comparative lack of ionization. Because
the internal energy of the gas is not stored as hydrogen ionization
energy, it remains as kinetic energy of the gas particles, raising the
temperature
  \citep{1992ApJ...397L..59C}. 

The temperatures also differ greatly in the cool intershock
regions. In LTE, they are between~3,000 and 5,000~K whereas the
non-equilibrium simulation reaches intershock temperatures of about
2,500~K.  These low values result from the reverse process:
over-ionization compared with LTE. More energy remains stored as
ionization energy, leaving less kinetic energy for the gas particles.

\section{Discussion \& Conclusions}

\subsection{Limitations of the simulation}

Our implementation of non-equilibrium hydrogen ionization has various
limitations.

First, the assumption that all Lyman transitions are in detailed
balance is justified up to the transition region  
  \citep{sollum1999}.
However, the transition region is optically thin in most Lyman
features, requiring detailed radiative transfer modeling to evaluate
their influence on the hydrogen populations.

Second, the multi-group radiative transfer within the simulation,
which sets the radiative cooling and heating, employs LTE ionization.
For given internal energy and mass density, the radiative transfer
uses the group-mean opacity, scattering probability and Planck
function based on the corresponding LTE (or coronal equilibrium)
temperature and electron density.  The radiative cooling in the
chromosphere and transition region, where deviations from equilibrium
are largest, is thus inconsistent with the non-equilibrium temperature
and electron density as computed in the simulation.

Third, the cool parts of the simulation chromosphere often reach the
limiting temperature of 2,400~K allowed in the simulation.  It is not
clear how low the actual chromospheric minima may reach because
radiative heating in the hydrogen continua and other strong spectral
features is not taken into account in the simulation, only their
radiative cooling.

\subsection{Discussion}

From the analysis of our simulation we obtain the following
picture. The internetwork chromosphere is irregularly pervaded by
shocks. The temperature is typically 10,000~K in shocks and can be
lower than 2,500~K in cool intershock areas. The transition region is
arc-shaped, with the arc footpoints situated above the magnetic
elements, and reaches a maximum height of 4~Mm. The chromosphere above
magnetic elements shows upward propagating shocks at about 3~minute
periodicity, which push the corona upward from 1.5~Mm to 3~Mm height.

The chromospheric hydrogen ionization degree increases smoothly from
$10^{-5}$ at 0.5~Mm to complete ionization in the corona and does not
respond to temperature changes because of the slow recombination
behind shocks. The hydrogen $n \is 2$ population in the chromosphere
is coupled to the proton population and, as a consequence, varies only
weakly with time too. The $n \is 2$ population in dynamic fibrils
above the magnetic elements is higher than in the internetwork
chromosphere. The $n \is 2$ column density, a measure of the optical
depth of the \Halpha\ line, is two orders of magnitude larger in
dynamic fibrils than at equal height in the internetwork. The same
column density as at the top of the fibrils is only reached at a
smaller height, about 0.8~Mm, in the internetwork.

In the observations of 
\citet{2006ApJ...647L..73H}
 and
\citet{2007ApJ...655..624D}  
dynamic fibrils appear as dark ``fingers'' extending and retracting on
top of a brighter background. They seem to be optically
thick. Combining this observational appearance with the results of our
simulation suggests that dynamic fibrils are optically thick in the
\Halpha\ line core, whereas optical depth unity is reached far deeper
in the internetwork atmosphere (equal column density in the lower
panels of Fig.~\ref{fig:ztslice}) so that the internetwork
chromosphere adjacent to dynamic fibrils is optically thin.  This
explains that dynamic fibrils are observable along a slanted line of
sight through the internetwork chromosphere.  We also suggest that
their low temperature combined with the effects of NLTE resonance
scattering produces their low emergent intensity.  The bright
background is formed much deeper in the internetwork, where the higher
temperature leads to higher emergent intensity.  In summary, slow
recombination and strong coupling of the $n \is 2$ population to the
ion population makes these fibrils opaque and therefore visible even
when they are cool, regardless of the large $n \is 2$ excitation
energy.  \Halpha\ interpretation assuming a static atmosphere with
instantaneous ionization and recombination, as frequently done in
cloud modeling and inversions based on cloud modeling, is likely to be
erroneous for such dynamic structures.  Obviously, detailed radiative
transfer computation based on time-dependent simulation with
non-equilibrium ionization as done here is needed to properly assess
\Halpha\ formation in dynamic fibrils.

\subsection{Conclusions}

We have presented an algorithm to compute non-equilibrium hydrogen
ionization with back-coupling to the equation of state in
multidimensional radiation MHD simulations of the solar atmosphere. We
performed a 2D simulation from the convection zone to the corona that
employed this algorithm. From its analysis and comparison with a
companion LTE simulation we conclude the following:
\begin{itemize}
\vspace{-0.5ex}

  \item Inclusion of non-equilibrium hydrogen ionization is essential
        in simulations of the solar atmosphere because the resulting
        temperature structure and hydrogen populations differ
        dramatically from their LTE values.

  \item The degree of ionization of hydrogen in the chromosphere does
        not follow the local temperature, as described
        already by
 \citet{2002ApJ...572..626C} 
        and
 \citet{2006A&A...460..301L}. 
        Hydrogen is partially ionized in shocks but does not
        recombine in the cool shock wakes, owing to the slow
        recombination rate at low temperature
        (Figs.~\ref{fig:xyslice},~\ref{fig:columns},
        and~\ref{fig:ztslice}).

  \item Non-equilibrium hydrogen ionization causes more profound
    temperature variations in the chromosphere than would occur if LTE
    were valid (Fig.~\ref{fig:xtslice}). The shock temperatures are
    higher and the intershock temperatures are lower, caused by the
    insensitivity of the hydrogen ionization degree to variations of
    the state parameters of the gas.

  \item The population of the hydrogen $n \is 2$ level in the
    chromosphere is strongly coupled to the ion population.  It
    therefore behaves approximately as the latter.  Its value is set
    by the high shock temperatures and subsequently remains high in
    the cool aftershock phases (Fig.~\ref{fig:columns}).  This is
    quite contrary to the LTE prediction.

  \item The simulation shows large differences in temperature
    structure and hydrogen level populations between magnetic elements
    and internetwork (Figs.~\ref{fig:columns} and~\ref{fig:ztslice}).

  \item The \Halpha\ line opacity is proportional to the $n \is 2$
    level population; the \Halpha\ optical depth scales with the $n
    \is 2$ column density.  Both are appreciably larger in dynamic
    fibrils than in the internetwork chromosphere at equal height
    (Fig.~\ref{fig:ztslice}). We suggest that dynamic fibrils are
    optically thick in \Halpha\ and that their low temperature
    combined with scattering make them appear dark against the
    deeper-formed bright internetwork background.

\end{itemize}

The next step is to compute \Halpha\ in detail from this simulation.

\begin{acknowledgements}
  J.~Leenaarts recognizes support from the
  USO-SP International Graduate School for Solar Physics (EU
  contract nr.~MEST-CT-2005-020395), and hospitality at the
  Institutt for Teoretisk Astrofysikk in Oslo.
\end{acknowledgements}

\bibliographystyle{aa} 
\bibliography{%
Carlsson,%
Freytag,%
Galsgaard,%
Hansteen,%
Johnson,%
Kneer,%
Leenaarts,%
Nordlund,%
Peter,%
Rammacher,%
Skartlien,%
Sollum,%
Stein,%
temp,%
Tsuji,%
Vardya}

\begin{appendix} \label{appendix:a}
\section{The equations and their derivatives}
The solution scheme requires partial derivatives with respect to the
independent variables.  The formulation is similar to the RADYN code
of
\citet{1992ApJ...397L..59C}
but since no detailed specification has been published so far, we
supply the complete derivative list here.

The present code solves the equations of chemical equilibrium, charge,
energy and hydrogen nucleus conservation together with the rate
equations or with the Saha-Boltzmann equilibrium equations.  These
equations depend on the gas temperature $T$, the electron density
\nel, the hydrogen population densities $n_i$, where $n_1$ is the
ground state, $n_{2-5}$ the four excited states and $n_6$ the ionized
hydrogen (proton) density, and the molecular hydrogen density
$\nh2$. These are ten equations and nine unknowns.  We use only five
of the six rate equations and eliminate the molecular hydrogen
density, reducing the number of equations and variables to eight. The
equations are labeled $F_i$ with $i=1,\ldots,8$.

The physical constants have their usual meaning: $h$, $k$, $c$, $e$ and
\me\ are Planck's constant, Boltzmann's constant, the velocity of
light, the electron charge and the electron mass.

As the mass density of the gas is known, the total number of hydrogen
nuclei, whether in the form of H, H$_2$ or bare protons is known and
given by
\begin{equation}
\ntot \is \rho \, n_\mathrm{H\, per\, g}
\end{equation} 
where $n_\mathrm{H\, per\, g}$ is the number of hydrogen nuclei per
gram solar gas. We denote the number density of nuclei of elements
other than hydrogen as $\nother= n_\mathrm{nuclei\, per\, H} \, \ntot$,
where $n_\mathrm{nuclei\, per\, H}$ is the number of nuclei from
elements other than hydrogen per hydrogen nucleus.  The number of
electrons due to elements other than hydrogen per hydrogen nucleus is
$\neother$. The internal energy of elements other than hydrogen per
hydrogen nucleus is $\eother$. These quantities and their derivatives
are determined numerically from tables as function of $T$ and
\nel. The molecular hydrogen density is \nh2 and the internal energy
of H$_2$ per molecule is $\eh2$. The latter is computed using
polynomial approximations by
 \citet{1965MNRAS.129..205V} 
%

\paragraph{Chemical equilibrium.}
The equation of chemical equilibrium between neutral atomic hydrogen
and molecular hydrogen is given by
\begin{equation}
  \nh2 = \frac{\left(\sum_{i=1}^5 n_i \right)^2}{K},
\end{equation}
where $K$ is the chemical equilibrium constant whose value and its
temperature derivative are given by polynomial approximations as
function of $T$ by
 \citet{1973A&A....23..411T}.
%
The molecular hydrogen density depends on $T$ and $n_i$ ($i=1,\ldots,5$). The
derivatives are
\begin{eqnarray}
  \frac{\partial \nh2}{\partial T}  & = &   
  - \left( \frac{\sum_{i=1}^5 n_i}{K} \right)^2
  \frac{\partial K}{\partial T}, \\
  \frac{\partial \nh2}{\partial n_i}  & = &  
  \frac{ 2 \left( \sum_{i=1}^5 n_i \right)  }{K}.
\end{eqnarray}

\paragraph{Charge conservation.}
The electron density is given by
\begin{equation}
\nel =  n_6 + \ntot \neother,
\end{equation}
which we rewrite in a form suitable for Newton-Raphson iteration:
\begin{equation}
F_1 = 1 - \frac{1}{\nel} \left( n_6 + \ntot \neother \right) = 0.
\end{equation}
The functional $F_1$ depends on $\nel$, $n_6$ and $T$. The partial
derivatives are
\begin{eqnarray}
  \frac{\partial F_1}{\partial \nel} & = & \frac{n_6+\ntot
    \neother}{\nel^2}- \frac{\ntot}{\nel} \frac{\partial
    \neother}{\partial \nel},\\ 
  \frac{\partial F_1}{\partial n_6} & = & -\frac{1}{\nel}, \\ 
  \frac{\partial F_1}{\partial T} & = &
    -\frac{\ntot}{\nel} \frac{\partial \neother}{\partial T}.
\end{eqnarray}

\paragraph{Internal energy.}
The second functional specifies the distribution of the internal energy
\ei\ over the various contributions:
\begin{eqnarray}
F_2 & = & 1 - \frac{1}{\ei} \left( \frac{3 k T}{2} \left[\nel  + \nother +
\nh2 +  \sum_{i=1}^6 n_i \right] \nonumber \right. \\
& & + \left. \ntot \, \eother + \nh2 \, \eh2 + \sum_{i=1}^6 n_i \, \chi_i \right) =0,
\end{eqnarray}
with $\chi_i$ the energy of hydrogen level $i$. This functional
depends on $\nel$, $n_i$ and $T$. The partial derivatives are
\begin{eqnarray}     
  \frac{\partial F_2}{\partial \nel} & = & -\frac{1}{\ei} \left( 
    \frac{3 k T}{2} + \ntot \frac{\partial \eother}{\partial \nel} \right), \\ 
  \frac{\partial F_2}{\partial n_i} & = & -\frac{1}{\ei} \left( \frac{3 k T}{2} 
     \left[ \frac{\partial \nh2}{\partial n_i} + 1 \right] +
      \eh2 \frac{\partial \nh2}{\partial n_i} + \chi_i \right), \\
  \frac{\partial F_2}{\partial T} & = & - \frac{1}{\ei} \left( \frac{3 k}{2}  
      \left[ \nel + \nother + \nh2  + T \frac{\partial \nh2}{\partial T} + \sum_{i=1}^6 n_i 
       \right] \right. \nonumber \\
 & & \left.  + \ntot \frac{\partial \eother}{\partial T} + \nh2 \frac{\partial \eh2}{\partial T} + \eh2
      \frac{\partial \nh2}{\partial T}  \right). 
\end{eqnarray}

\paragraph{Particle conservation.}
The conservation of the total number of hydrogen nuclei $\ntot$ is given by:
\begin{equation}
   F_3 = 1 - \frac{1}{\ntot} \left( \sum_{i=1}^6 n_i - 2 \nh2 \right) = 0.
\end{equation}
This functional depends on $n_i$ and $T$. The partial derivatives are:
\begin{eqnarray}
  \frac{\partial F_3}{\partial n_i}  &= &   -\frac{1}{\ntot} 
      \left( 1 + 2  \frac{\partial \nh2}{\partial n_i}   \right), i=1,\ldots,5, \\
  \frac{\partial F_3}{\partial n_6} & =  &  -\frac{1}{\ntot}, \\
      \frac{\partial F_3}{\partial T} & =  &  - \frac{2}{\ntot} \frac{\partial \nh2}{\partial T}.
\end{eqnarray}

\paragraph{LTE populations.} 
LTE is imposed to start up a new computation and at the lower
boundary. The LTE hydrogen populations then obey the Saha-Boltzmann
equations ($i=1,\ldots,5$)
\begin{equation}
  F_{3+i} = 1 - \frac{1}{\nel} \frac{n_i}{n_6} \frac{2 g_6}{g_i} \left(
  \frac{2 \pi \me k T}{h^2}\right)^{3/2} \exp^{-(\chi_6-\chi_i) / k T} =
  0,
 \end{equation}
where $g_i$ is the statistical weight of level $i$. These equations
depend on $n_6$, $n_i$, $\nel$ and $T$. Defining
\begin{equation}
 K_i =  \frac{1}{\nel} \frac{n_i}{n_6} \frac{2 g_6}{g_i} \left(
  \frac{2 \pi \me k T}{h^2}\right)^{3/2} \exp^{-(\chi_6-\chi_i) / k T},
\end{equation}
the partial derivatives are  given by
\begin{eqnarray}
  \frac{\partial F_{3+i}}{\partial n_6} & = & \frac{K_i}{n_6}, \\
  \frac{\partial F_{3+i}}{\partial n_i} & = & - \frac{K_i}{n_i}, \\
  \frac{\partial F_{3+i}}{\partial \nel} & = &  \frac{K_i}{\nel}, \\
  \frac{\partial F_{3+i}}{\partial T} & = & - K_i 
    \left(  \ \frac{\chi_6 - \chi_i}{k T^2}+\frac{3}{2T}  \right).
\end{eqnarray}

\paragraph{Rate equations.}

Outside LTE, the change of the hydrogen populations $n_1 \ldots n_6$
over a timestep $\Delta t$ can be expressed as
\begin{eqnarray}
  n_i(t_0+\Delta t) - n_i(t_0) &=& \int_{t_{0}}^{t_{0}+\Delta t} 
     \sum_{j,j \ne i}^6 n_j 
  \left(R_{ji} + C_{ji} \right) \nonumber \\
& &  - n_i \sum_{j,j \ne i}^6\left(R_{ij} + C_{ij} \right) \, \rmd t,
\label{eq:rateintegral}
\end{eqnarray}
with $R_{ij}$ and $C_{ij}$ the radiative and collisional rate
coefficients from level $i$ to level $j$.  Defining
\begin{eqnarray}
  P_{ij} & = & R_{ij} + C_{ij}, \\ 
  P_{ii} & = & - \sum_{j,j \ne i}^6\left(R_{ij}+C_{ij}\right)
\end{eqnarray} 
and writing $n_i = n_i\,(t_0+\Delta t)$ and $n_i^{t_0} = n_i\,(t_0)$
yields the discretized, implicit form of Eq.~\ref{eq:rateintegral}
\begin{equation}
 F_{3+i} = \frac{n_i}{n_i^{t_0} } - \frac{\Delta t}{n_i^{t_0}} 
   \left( \sum_{j=1}^6 n_j P_{ji} \right) - 1 = 0.
 \end{equation}
The equations are dependent on $n_i$, $n_j$ ($j \ne i$), $\nel$ and
$T$. The partial derivatives are
\begin{eqnarray}
  \frac{\partial F_{3+i}}{\partial n_i} & = & \frac{1-\Delta t \, P_{ii}}{n_i^{t_0}}, \\
  \frac{\partial F_{3+i}}{\partial n_j} & = &  -\frac{\Delta t \, P_{ji}}{n_i^{t_0}}, \\
  \frac{\partial F_{3+i}}{\partial \nel} & = & - \frac{\Delta t}{n_i^{t_0}} 
     \left( \sum_{j=1}^6 n_j \frac{ \partial P_{ji}}{\partial \nel} \right), \\        
  \frac{\partial F_{3+i}}{\partial T} & = &  - \frac{\Delta t}{n_i^{t_0}}  
    \left( \sum_{j=1}^6 n_j \frac{ \partial P_{ji}}{\partial T} \right).                      
\end{eqnarray}
Only five of the six equations~\ref{eq:rateintegral} are used to avoid
overspecification through ignoring the one with the largest value of
$n_i^{t_0}$.

\paragraph{LTE population ratios.}
Evaluation of the rate coefficients involving the continuum level
$n_6$ often involve the LTE population ratio $n^*_i/n^*_6$ ($i=1,\ldots,5$) given by
\begin{equation}
   \frac{n^*_i}{n^*_6}= \nel \frac{g_i}{2 g_6} \left( \frac{2 \pi \me k
     T}{h^2}\right)^{-3/2} \exp^{(\chi_6-\chi_i) / k T},
\end{equation}
with the derivatives
\begin{eqnarray}
\frac{\partial}{\partial T}  \left( \frac{n^*_6}{n^*_i} \right) &  = &  
- \left(  \ \frac{\chi_6 - \chi_i}{k T^2}+\frac{3}{2T}  \right) \frac{n^*_i}{n^*_6}, \\
 \frac{\partial}{\partial \nel}  \left( \frac{n^*_i}{n^*_6} \right)  
 &  = &  \frac{1}{\nel}  \frac{n^*_i}{n^*_6}.  
\end{eqnarray}

\paragraph{Collisional rate coefficients.}
 The expressions for the collisional rate coefficients contain
 temperature-dependent coefficients $C_\mathrm{exc}$ and
 $C_\mathrm{ion}$ for excitation and ionization. Their values and
 derivatives are determined from a table based on
  \citet{1972ApJ...174..227J}.

The collisional rate coefficients for bound-bound transitions are, with $j>
 i$ and $C_{ij}$ the upward rate coefficient:
\begin{eqnarray}
 C_{ji} &  = & \frac{g_i}{g_j} \nel C_\mathrm{exc} \sqrt{T}, \\
 C_{ij} &  = & \frac{n^*_j}{n^*_i} C_{ji} =  \nel   C_\mathrm{exc} \sqrt{T} \, \exp^{-(\chi_j-\chi_i)/k T},
\end{eqnarray}
with derivatives
\begin{eqnarray}
   \frac{\partial C_{ji}}{\partial \nel } &  = & \frac{C_{ji}}{\nel}, \\
   \frac{\partial C_{ji}}{\partial T} &  = &  \frac{g_i}{g_j} \nel \sqrt{T}  
      \left( \frac{\partial C_\mathrm{exc}}{\partial T} + \frac{C_\mathrm{exc}}{2T}   \right), \\
   \frac{\partial C_{ij}}{\partial \nel } &  = & \frac{C_{ij}}{\nel}, \\
   \frac{\partial C_{ij}}{\partial T} &  = & \nel \sqrt{T}  \exp^{-(\chi_j-\chi_i)/k T} 
      \left( \frac{\partial  C_\mathrm{exc}}{\partial T} +  C_\mathrm{exc} 
      \left[ \frac{\chi_j-\chi_i}{k T^2}+\frac{1}{2T}\right] \right)
\end{eqnarray}

For bound-free transitions the collision rate coefficients are,
with $C_{i6}$ the upward rate coefficient:
\begin{eqnarray}
   C_{i6} &  = & \nel C_\mathrm{ion}(T) \sqrt{T} \exp^{-(\chi_6-\chi_i)/k T},\\
   C_{6i} &  = & \frac{n^*_i}{n^*_6} C_{i6}, 
\end{eqnarray}
with derivatives
\begin{eqnarray}
  \frac{\partial C_{i6}}{\partial \nel } &  = & \frac{C_{i6}}{\nel}, \\
  \frac{\partial C_{i6}}{\partial T} &  = & \nel \sqrt{T}  
    \exp^{-(\chi_6-\chi_i)/k T} \left( \frac{\partial C_\mathrm{ion}}{\partial T} + 
    C_\mathrm{ion} \left[ \frac{\chi_6-\chi_i}{k T^2}+\frac{1}{2T}\right] \right) \\
  \frac{\partial C_{6i}}{\partial \nel } &  = & 2 \frac{C_{6i}}{\nel}, \\
  \frac{\partial C_{6i}}{\partial T} &  = & 
    \left( \frac{\partial C_{i6}}{\partial T} - C_{6i} \left[  \frac{3}{2T} 
    + \frac{\chi_6 - \chi_i}{k T^2}  \right] \right)  \frac{n^*_6}{n^*_i}.
\end{eqnarray}

\paragraph{Radiative rate coefficients.}

The derivation of the radiative rate coefficients can be found in
 \citet{sollum1999}.
The rate coefficients for bound-bound transitions are, with $j> i$ and
$R_{ij}$ the upward rate coefficient:
\begin{eqnarray}
 R_{ij} &  = &   \frac{4 \pi^2 e^2}{h \nu_0 \me c} f_{\rml \rmu}  
 \frac{2 h \nu_0^3}{c^2} \frac{1}{\rme^{h \nu_0 /  k \Trad} - 1},  \\
 R_{ji} & = & \frac{g_i}{g_j} \frac{4 \pi^2 e^2}{h \nu_0 \me c} 
 f_{\rml \rmu}  \frac{2 h \nu_0^3}{c^2} \frac{1}{1 - \rme^{-h \nu_0 /  k \Trad} },
\end{eqnarray}
where $f_{\rml \rmu}$, $\nu_0$ and $\Trad$ are the oscillator
strength, the line center frequency and the prescribed radiation
temperature. In areas where $\Trad \is T$, \ie\ deep in the
atmosphere, the temperature derivatives are:
\begin{eqnarray}
 \frac{\partial R_{ij}}{\partial T} &  = &     
   \frac{8 \pi^2 e^2 h \nu_0^3}{\me c^3 k T^2}f_{ij}
   \frac{\exp^{-h \nu_0 /  k T}}{(1-\exp^{-h \nu_0 /  k T} )^2},  \\  
 \frac{\partial R_{ji}}{\partial T} &  = & \frac{g_i}{g_j}  
   \frac{8 \pi^2 e^2 h \nu_0^3}{\me c^3 k T^2}f_{ij}  
   \frac{\exp^{h \nu_0 /  k T}}{(\exp^{h \nu_0 /  k T}-1 )^2},
\end{eqnarray}
In areas where $\Trad \ne T$, but constant, the temperature derivatives
are zero.

The upward radiative rate coefficient for bound-free transitions is:
\begin{equation}
 R_{i6} = \frac{8 \pi}{c^2} \alpha_0 \nu_0^3 \sum_{n=1}^\infty
   E_1 \left( n \frac{h \nu_0} {k \Trad} \right),
\end{equation}
where $\alpha_0$ is the radiative absorption cross section at the
ionization edge frequency $\nu_0$ and $E_1(x)$ the first exponential
integral with argument $x$. Its temperature derivative is:
\begin{eqnarray}
  \frac{\partial R_{i6}}{\partial T}  & = & \frac{8 \pi}{c^2} \alpha_0 
  \nu_0^3 \frac{1}{T} \frac{1}{\exp^{h \nu_0/k T} -1}\ \mathrm{where}\ T_\mathrm{rad} \is T, \\
 \frac{\partial R_{i6}}{\partial T}  & = & 0\ \mathrm{where}\ T_\mathrm{rad} \ne T.
\end{eqnarray}

The downward bound-free radiative rate coefficient is:
\begin{equation}
 R_{6i}  =   \frac{8 \pi}{c^2} \alpha_0 \nu_0^3 \frac{n^*_i}{n^*_6}
   \sum_{n=1}^\infty E_1 \left( \left[ n \frac{T}{\Trad} +1 \right] 
   \frac{h \nu_0}{k T} \right),
\end{equation}
 The derivatives are:
\begin{eqnarray}
  \frac{\partial R_{6i}}{\partial \nel}  & = &\frac{R_{6i}}{\nel}, \\
  \frac{\partial R_{6i}}{\partial T}  & = & \frac{\partial R_{i6}}{\partial T} \frac{n^*_i}{n^*_6} - R_{6i} 
    \left( \frac{3}{2T} + \frac{h \nu_0}{k T^2}\right)\ \mathrm{where}\ T_\mathrm{rad} \is T, \\
    \frac{\partial R_{6i}}{\partial T} & = & \frac{8 \pi}{c^2} \alpha_0 
    \nu_0^3 \left( \sum_{n=1}^\infty \frac{\mathrm{exp}\left(  - n
    \frac{h \nu_0}{k \Trad} - \frac{h \nu_0}{k T} \right)}{n T +
    \Trad} \frac{\Trad}{T} \right) \frac{n^*_i}{n^*_6} \nonumber \\
    &  & - R_{6i} \left( \frac{3}{2T} + \frac{h \nu_0}{k T^2}\right)\ \mathrm{where}\ T_\mathrm{rad} \ne T.
\end{eqnarray}

\end{appendix}

\end{document}